\documentclass[preprint,showpacs,preprintnumbers,amsmath,amssymb]{revtex4}

\usepackage{graphicx}% Include figure files
\usepackage{dcolumn}% Align table columns on decimal point
\usepackage{bm}% bold math
\begin{document}

\title{On the density dependence of single-proton and two-proton 
  knockout reactions under quasifree conditions}

\author{Wim Cosyn}
\email{Wim.Cosyn@UGent.be}
\author{Jan Ryckebusch}
\email{Jan.Ryckebusch@UGent.be}

\affiliation{Department of Subatomic and Radiation Physics,\\
 Ghent University, Proeftuinstraat 86, B-9000 Gent, Belgium}
\date{\today}

\begin{abstract}
We consider high-energy quasifree single- and two-proton knockout
reactions induced by electrons and protons and address the question
what target-nucleus densities can be effectively probed after
correcting for nuclear attenuation (initial- and final-state
interactions). Our calculations refer to ejected proton kinetic
energies of 1.5 GeV, the reactions $(e,e'p)$, $(\gamma,pp)$ and
$(p,2p)$ and a carbon target.  It is shown that each of the three
reactions is characterized by a distinctive sensitivity to the density
of the target nucleus. The bulk of the $(\gamma,pp)$ strength stems
from the high-density regions in the deep nuclear interior.  Despite
the strong attenuation, sizable densities can be probed by $(p,2p)$
provided that the energy resolution allows one to pick nucleons from
$s$ orbits.  The effective mean densities that can be probed in
high-energy $(e,e'p)$ are of the order of 30-50\% of the nuclear
saturation density.
\end{abstract}

%24. 	Nuclear reactions: general
% 
%24.10.-i 	Nuclear reaction models and methods
%24.10.Jv 	Relativistic models
%25.30.Rw 	Electroproduction reactions
%25.40.Ep 	Inelastic proton scattering


\maketitle

Nucleon knockout studies in quasifree kinematics belong to the most
powerful instruments for studying the structure of nuclei. Since the
1960's the $A(e,e'p)$ reaction has provided a wealth of information
about the merits and limitations of the nuclear shell-model
\cite{ISI:A1996BF20A00002}.  Quasifree proton scattering from nuclei
$A(p,2p)$ has a somewhat longer history \cite{RevModPhys.45.6} and
could in principle provide similar information as $A(e,e'p)$. With
three protons subject to attenuation effects, in $A(p,2p)$ the
description of the initial and final-state interactions (ISIs and
FSIs), is a more challenging issue than in $A(e,e'p)$.  Recent
applications of $A(p,2p)$ involve the analyzing power $(A_y)$ as an
instrument for probing medium modifications of hadron properties and
the density dependence of the nucleon-nucleon interaction
\cite{PhysRevLett.78.1014} \cite{PhysRevC.69.024604}. In inverse
kinematics (i.e. the $p(A,2p)A-1$ reaction) the $(p,2p)$ process
offers great opportunities for systematic studies of the density and
isospin dependency of single-particle properties in unstable nuclei
\cite{Kobayashi2008} at high-energy radioactive beam facilities
\cite{Aumann2007}. Experiments of that type have the potential to
study the equation-of-state for nuclei far from equilibrium. With
regard to quasifree $A(e,e'p)$, recent developments include the search
for medium modifications of electromagnetic form factors through
double polarization experiments of the type $^{4}$He$(\vec{e},e'
\vec{p})$ \cite{PhysRevLett.91.052301}. Another line of current
research is that of the two-nucleon removal reactions $A(e,e'pp)$ and
$A(\gamma,pp)$ in selected kinematics.  These reactions provide a
window on the short-range structure of nuclei \cite{shneor:072501}.

The development of an appropriate reaction theory is essential for
reliably extracting the physical information from the nucleon knockout
reactions.  For nucleon kinetic energies up to about 1 GeV the
distorted wave impulse approximation (DWIA) has enjoyed many successes
in that it could reproduce a large amount of measurements fairly well
\cite{ISI:A1996BF20A00002}. Constraining the optical potentials for
DWIA calculations, however, heavily depends on the availability of
elastic proton-nucleus scattering data. Moreover, the optical
potentials exhibit a substantial kinetic-energy dependence. This
energy dependence makes it difficult to make more general statements
about e.g. the role of attenuation effects and the effective densities
that can be probed in the various reactions. At sufficiently high
nucleon energies the Glauber approach provides a valid and highly
efficient alternative for the DWIA framework.  The Glauber approach
has the advantage that the effect of ISIs and FSIs can be computed
from the knowledge of the elementary proton-proton and proton-neutron
differential cross sections and of the density of the target (residual)
nucleus. Moreover, for nucleon momenta exceeding about 1 GeV, the
energy dependence of the parameters entering the Glauber calculations
is relatively smooth. This results, for example, in measured and
computed nuclear $A(e,e'p)$ transparencies which exhibit little energy
dependence nucleon kinetic energies larger than 0.5~GeV
\cite{Lava:2004zi}. From the theoretical point of view, it allows one
to make more universal statements about the predicted role of nuclear
attenuation.  Another advantage of the Glauber approach is that it is
applicable to a wide range of reactions, including electromagnetic and
hadronic probes, with stable and unstable nuclei
\cite{ISI:000255457700024} \cite{Hussein:1990fr}.

We exploit the robustness of the Glauber approach to
study the density dependence of quasifree nucleon removal
reactions. Indeed, investigations into the medium dependence of
nucleon properties and the study of the nuclear structure of unstable
nuclei e.g., heavily rely on the possibility of effectively probing
regions of sufficiently high density in the target nucleus.  Nuclear
attenuation effects on the impinging and ejected protons can cause 
the nucleon knockout reactions to effectively probe regions of relatively
small density near the surface of the target nucleus.  The description
of nuclear attenuation brings in a certain
degree of model dependence. We stress the importance of making cross
checks over different types of reactions (electromagnetic versus
hadronic probes) and linking single-nucleon to two-nucleon knockout 
reactions. Here, we report on a study of the effective nuclear density
that can be probed in reactions that have one nucleon ($A(e,e'p)$),
two nucleons ($A(\gamma,pp)$) and three nucleons ($A(p,2p)$) subject to
nuclear attenuation effects.

In Ref.~\cite{Ryckebusch:2003fc} a relativistic extension of the
Glauber method was introduced. The method was coined RMSGA
(relativistic multiple-scattering Glauber approximation). In line with
the assumptions of a typical relativistic DWIA model, the RMSGA model
uses a relativistic mean-field to describe the target and residual
nucleus in combination with the impulse approximation for the
interaction Hamiltonian. The RMSGA differs from the relativistic DWIA
in that it uses a relativistic extension of the Glauber method to
treat initial-state and final-state interactions.

\begin{figure*}
%\begin{figure}[htb]
\centering
\includegraphics[width=15.6cm]{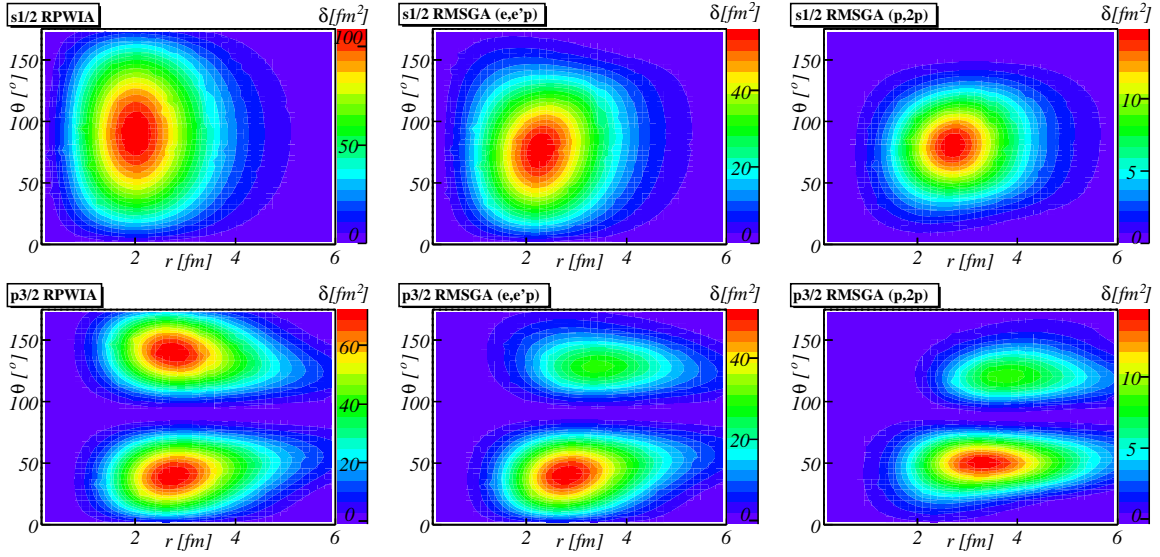}
\caption{\textit{(color online)} The function $\delta (r, \theta) $
  for the $^{12}$C$(e,e'p)$ and $^{12}$C$(p,2p)$ reaction. For both
  types of reactions we consider an energy transfer of 1.5 GeV and a
  three-momentum transfer $\vec{q}$ that is tuned to probe the maximum
  of the momentum distribution (i.e. $p_m$=0 MeV for knockout from the
  $s1/2$-orbit and $p_m$=115 MeV for removal from the $p3/2$-orbit).
  For the $(e,e'p)$ results the proton is detected along the direction
  of the momentum transfer. For the $(p,2p)$ the incoming proton has a
  kinetic energy of about 3 GeV and the two ejected protons have a
  kinetic energy of 1.5 GeV. They are detected under an angle of about
  32$^{o}$ but on opposite sides of the incoming beam. For the
  sake of reference, the proton root-mean-square radius in $^{12}$C as
  determined from elastic electron scattering is $\left< r^2 \right>
  ^{1/2} = 2.464 \pm 0.012 $~fm \cite{1982PhRvC..26..806R}.}
\label{figeep}
%\end{figure}
\end{figure*}

In a factorized approach, the differential cross sections for the
single-nucleon removal reactions considered here (i.e. $A(p,2p)$ and
$A(e,e'p)$) are proportional to the distorted momentum distributions
$ \rho _{\left( n \kappa \right)} ^{D} (\vec{p}_m) $
\begin{eqnarray}
\rho _{ n \kappa  } ^{D} (\vec{p}_{miss}) & = &
\sum_{s , m}
\left|  \int d \vec{r}
\frac { 
 e^{-i\vec{p} _{m} \cdot \vec{r}}
} 
{(2\pi)^{3}} 
\bar{u}(\vec{p}_m, s) 
\widehat{\mathcal{S}}_{\text{RMSGA}}^\dagger(\vec{r})
\phi_{ n \kappa m } (\vec{r})
\right|^2 \; , 
\nonumber \\
& = & \sum _{s, m} \left( {\phi_ {n \kappa m }  ^{D} } \right) ^{\dagger} 
{\phi_ {n \kappa m }  ^{D} }
\; ,
\nonumber \\
& = & \frac {1} {2} \int d r \int d \theta 
\left[ \sum _{s,m} \left( \left( { D(r, \theta) } \right) ^{\dagger} {\phi_ {n 
      \kappa m  }
    ^D } + 
 { D(r, \theta) } \left( {\phi_ { n \kappa m  } ^D } \right)
 ^{\dagger} \right) \right] \; ,   
\nonumber \\
& \equiv & \int d r \int d \theta \delta \left( r, \theta \right) \; 
\; ,
\label{eq:rhormsga}
\end{eqnarray}
where the quantum numbers $(n \kappa m) $ determine the orbit
of the struck nucleon, $ \phi_{ n \kappa m }
(\vec{r}) $ is the corresponding relativistic single-particle wave
function and $u(\vec{p},s)$ a four-component free-particle spinor.
The missing momentum $\vec{p}_{m}$ is determined by the difference
between the asymptotic three-momentum  of the ejected nucleon
$\vec{p}$ 
and the three-momentum transfer $ \vec{q}$. We define the $z$-axis
along the $\vec{q}$ and the $xz$-plane as the reaction plane.  The
function $\delta ( r , \theta)$ defined in Eq.~(\ref{eq:rhormsga})
encodes the contribution from an infinitesimal interval around $r$ and
$\theta$ to a single-nucleon removal cross section
\cite{PhysRevLett.78.1014}.  The function $D(r, \theta)$ which was
introduced in (\ref{eq:rhormsga}) reads
\begin{equation}
D(r, \theta) = \int d \phi \; r^2 \; \sin \theta \frac { 
 e^{-i\vec{p} _{m} \cdot \vec{r}}
} 
{(2\pi)^{3}} 
\bar{u}(\vec{p}_m, s) 
\widehat{\mathcal{S}}_{\text{RMSGA}}^\dagger(\vec{r})
\phi_{ n \kappa m } (\vec{r}) \; .
 \end{equation}
The Glauber phase
operator $ \widehat{\mathcal{S}}_{\text{RMSGA}}^\dagger(\vec{r}) $
encodes the combined effect of the initial and final-state
interactions.  Retaining only central interactions, which is a fair
approximation at the kinetic energies considered here, it can be
considered as a scalar operator. For the $A(p,2p)$ reaction, the $
\widehat{\mathcal{S}}_{\text{RMSGA}} $ becomes a multi-dimensional
convolution over the squared wavefunctions of the spectator nucleons
times the profile functions for the impinging proton and two ejected
protons \cite{overmeire:064603}. For the $A(e,e'p)$ case the
convolution involves the squared wavefunctions of the spectator
nucleons times the profile function for the ejected nucleon
\cite{Ryckebusch:2003fc}.

\begin{figure}[htb]
\centering
\includegraphics[width=8.6cm]{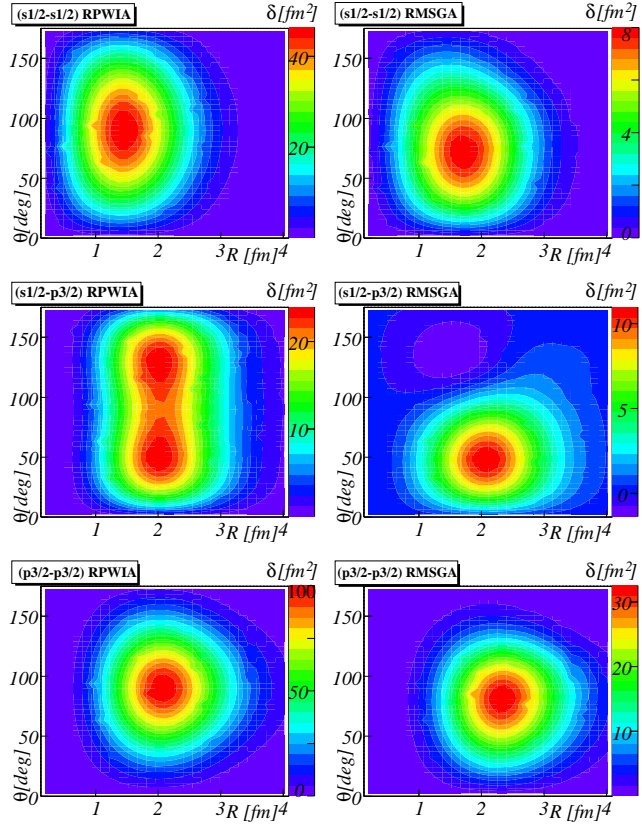}
\caption{\textit{(color online)} The function $\delta (R, \theta) $
  for the exclusive $^{12}$C$(\gamma, pp)$ cross section. In all
  situations we consider an energy transfer of 3 GeV and a
  three-momentum transfer $\vec{q}$ that is tuned to probe the maximum
  of the momentum distribution $\rho _{n_1 \kappa _1, n_2 \kappa _2} (\vec{P})
  $ (i.e. $P$=0 MeV for knockout from the $(s1/2 -s1/2)$- and $(p3/2 -
  p3/2)$-orbits and $P$=160 MeV for removal from the $(s1/2 -
  p3/2)$-orbits). We consider coplanar and symmetric kinematics, i.e.
  the two escaping protons have the same energy and polar angle
  $\theta_{pq}$, but escape from the opposite side of $\vec{q}$.}
\label{figeepp}
\end{figure}

We now wish to formulate the analog of Eq.~(\ref{eq:rhormsga}) for
two-nucleon knockout reactions of the type $A(e,e'pp)$ and
$A(\gamma,pp)$.  The asymptotic three-momenta of the two ejected
protons are defined as $\vec{p}_1$ and $\vec{p}_2$.  We introduce
relative $\vec{p} = \frac {\vec{p}_1 - \vec{p}_2} {2} $ and
center-of-mass $\vec{P} = \vec{p}_1 + \vec{p}_2 $ momenta of the
ejected pair. Factorizing the $A(\gamma ,pp)$ and $A(e,e'pp)$ cross
sections requires the assumption that the sudden emission of two
correlated protons only occurs when they reside in a relative $S$
state \cite{Ryckebusch:1996wc}. This is a reasonable approximation as
investigations of the $^{16}$O$(e,e'pp)$ reaction at the electron
accelerators in Mainz \cite{ISI:000222562800013} and Amsterdam
\cite{ISI:000085387800006} have shown that proton pairs are
exclusively subject to short-range correlations when they reside in a
relative $S$ state under conditions corresponding with relatively
small center-of-mass momenta $P$. In other words, the distinctive
feature of the correlated two protons is that they are very close and
moving back-to-back.  With the relative $S$-state assumption, the
differential cross section is proportional to the distorted momentum
distribution corresponding with the center-of-mass motion of the pair
%\begin{widetext}
\begin{eqnarray}
\rho _{n_1 {\kappa}_1, n_2 {\kappa}_2} ^{D} (\vec{P}) & = &
\sum_{s_1, s_2, m_1, m_2}
\left|  \int d \vec{R}
\frac { 
 e^{-i\vec{P}_m \cdot \vec{R}}
} 
{(2\pi)^{3}} 
\bar{u}(\vec{p} + \frac {\vec{P}} {2}, s_1)
\phi_{n _{1} \kappa _1 m_1} (\vec{R})
\right.
\nonumber \\
& & \left. \times 
\bar{u}(-\vec{p} + \frac {\vec{P}} {2}, s_2) 
\phi_{n_{2} \kappa_2 m_2} (\vec{R})
\widehat{\mathcal{S}}_{\text{RMSGA}}^\dagger(\vec{R})
\right|^2 
\nonumber \\
& \equiv & \int d R \int d \theta \delta \left( R, \theta \right) \; 
\; ,
\label{eq:rho2rmsga}
\end{eqnarray}
%\end{widetext}
here, $\vec{P}_m = \vec{p}_1 + \vec{p}_2 - \vec{q}$ is the pair's
missing momentum in the quasifree approximation. It can be interpreted
as the center-of-mass momentum of the correlated proton pair that
absorbs the proton. The quantum numbers $ \left( n_1 {\kappa}_1, n_2
{\kappa}_2 \right)$ determine the orbits of the two correlated nucleons.

\begin{table}[htb]
\caption{The average effective density $\overline{\rho}$ probed in the
  various reactions.}
\begin{ruledtabular}
\begin{tabular}{|c|c|c|c|}
Reaction & orbits & $\overline{\rho}{(RPWIA)}$  &  $\overline{\rho}
(RMSGA)$  \\
& & (fm$^{-3}$) & (fm$^{-3}$)   \\
\hline 
$(p,2p)$ &  $s1/2$	& 0.100  & 0.055 \\
$(p,2p)$  &  $p3/2$ 	& 0.050  & 0.025  \\
$(e,e'p)$&  $s1/2$	& 0.100 & 0.086  \\
$(e,e'p)$ &  $p3/2$      & 0.050 & 0.038  \\
$(e,e'pp)$ & $(s1/2-s1/2)$	& 0.150  & 0.135  \\
$(e,e'pp)$ & $(p3/2-p3/2)$    & 0.095  & 0.075 \\
$(e,e'pp)$ & $(s1/2-p3/2)$    & 0.115  & 0.098  \\
\end{tabular}
\end{ruledtabular}
\label{tab:tab1}
\end{table}

We now present the results of the numerical calculations for $\delta
(r, \theta)$ and $ \delta (R, \theta) $. A relativistic
single-particle model was used for $^{12}$C with parameters adjusted
to describe the ground-state properties.  The profile functions
entering the $ \widehat{\mathcal{S}}_{\text{RMSGA}} $ operator have
three parameters that have been determined from the database of
proton-proton and proton-neutron cross sections
\cite{Ryckebusch:2003fc}.  In Fig.~\ref{figeep} we display the
function $ \delta (r,\theta)$ defined in the Eq.~(\ref{eq:rhormsga})
for proton knockout from the $s1/2$ and $p3/2$ orbit from a $^{12}$C
target.  We compare the $(p,2p)$ with the $(e,e'p)$ result for an
energy transfer of 1.5 GeV and conditions probing the maximum of the
undisturbed momentum distribution $\rho_{ n \kappa }
(\vec{p}) $.  The latter can be obtained by setting $
\widehat{\mathcal{S}}_{\text{RMSGA}} = 1 $ in
Eq.~(\ref{eq:rhormsga}). In the considered kinematics, it is clear
that in the absence of nuclear attenuation ($
\widehat{\mathcal{S}}_{\text{RMSGA}} = 1 $ ), the upper ($0 ^ {\circ}
\le \theta \le 90^ {\circ} $) and lower hemisphere ($90 ^ {\circ} \le
\theta \le 180^ {\circ} $) of the target nucleus equally contributes
to $\delta ( r , \theta)$ and the measured signal. Moreover, the
$\delta (r,\theta)$ becomes equal for $(e,e'p)$ and $(p,2p)$. We refer
to this situation as the relativistic plane-wave impulse approximation
(RPWIA). As the ISI and FSI have the strongest impact at the highest
nuclear densities, the RMSGA predictions for $\delta (r,\theta)$ are
shifted to larger values of $r$ in comparison with the RPWIA ones.  In
addition, the contribution from the upper and lower hemisphere becomes
asymmetric when considering attenuation. Indeed, the nuclear
hemisphere closest to the proton detector provides the strongest
contribution to the detected signal. The stronger the effect of
attenuation the larger $\delta (r, \theta)$ experiences a shifts in
$r$, the larger the induced asymmetries between the upper and lower
hemisphere and the stronger the reduction.  Obviously, the asymmetry,
shift and reduction occur for the $\delta (r,\theta) $ in $(e,e'p)$
and $(p,2p)$.  All three effects, however, are far more pronounced for
the $(p,2p)$ than for the corresponding $\delta (r, \theta)$ in $(e,e'p)$.

In Fig.~\ref{figeepp} we display for the $^{12}$C target the function
$\delta (R,\theta)$ defined in the Eq.~(\ref{eq:rho2rmsga}) for
two-proton knockout from the $(p3/2 - p3/2)$, $(s1/2 - s1/2)$ and
$(s1/2 - p3/2)$ orbits. Comparing Figs.~\ref{figeep} and
\ref{figeepp} it is clear that two-proton removal at high energies,
really succeeds in probing the high-density regions of the target
nucleus (note the different range in the radial coordinate $r$ for
Figs.~\ref{figeep} and \ref{figeepp}). The attenuation mechanisms
induce shifts to the surface but the bulk of the measured strength can
be clearly attributed to high-density regions in the target nucleus.

%\begin{figure}[htb]
%\centering
%\includegraphics[width=6.6cm]{radialdensity._fermi.eps}
%\caption{\textit{(color conline)} Contribution to the exclusive
%$^{12}$C$(e,e'p)$ (green) , $^{12}$C$(p,2p)$ (blue) and
%$^{12}$C$(e,e'pp)$ (blue) cross section as a function of the radial
%coordinate $r$.  The kinematic conditions are those of
%Figs.~\ref{figeep} and \ref{figeepp}. We consider one- and two-nucleon
%removal from the $p$-orbit.  The RPWIA result is displayed
%in red. The black solid line shows $r^2 \rho (r)$ for the $^{12}$C
%nucleus.}
%\label{figradial}
%\end{figure}

In order to quantify the average densities that the various reactions
can probe, we introduce \cite{PhysRevC.69.024604, PhysRevLett.78.1014}
\begin{equation}
\overline{\rho} = \frac 
{\int d r d \theta  \rho  \left( \vec{r} \right) \delta \left( r, \theta \right) }
{\int d r d \theta   \delta \left( r, \theta \right) } \; ,
\label{eq:averdens}
\end{equation}
where $\rho \left( \vec{r} \right)$ is the density of the target
nucleus and $ \delta \left( r, \theta \right) $ the function as it was
defined in Eq.~(\ref{eq:rhormsga}) (single-proton knockout) and
Eq.~(\ref{eq:rho2rmsga}) (two-proton knockout). Table \ref{tab:tab1}
lists a systematic comparison of the computed values of
$\overline{\rho}$.  The average density probed in the two-proton
removal reaction from the $(s1/2 - s1/2)$ orbits approaches the
nuclear saturation density of $\rho _0 = 0.17$~fm$^{-3}$.  We wish to
stress the strong dependence on the nuclear orbit. Despite the strong
attenuation, the $(p,2p)$ reaction from the $s1/2$ orbit can
effectively probe higher densities than the $(e,e'p)$ reaction from
the valence $p3/2$ shell.  For the $(p,2p)$ reaction with knockout
from the $s1/2$ orbit the predicted effective mean density from the
RMSGA calculations is $ \overline{\rho} \approx 0.33 ~\rho _0$.  This
number is almost identical to the DWIA results of
Ref.~\cite{PhysRevC.69.024604} for $^{12}$C$(p,2p)$ for 1 GeV incoming
protons. 

In summary, we have used a relativistic framework to make a
comparative and consistent study of the effective nuclear densities
that can be probed in $(p,2p)$, $(e,e'p)$ and $(\gamma,pp)$ reactions.
As a representative example we have selected a carbon target and
ejected proton kinetic energies of 1.5 GeV.  We consider the results
as representative for light nuclei and sufficiently high kinetic
energies.  The conclusions drawn in this work are of importance for
ongoing and planned searches of nuclear effects at small distance
scales. The $(e,e'p)$
reaction has the potential to probe reasonable densities.  Of all
reactions considered here, the $(\gamma, pp)$ reaction is the one that can
get closest to the deep nuclear interior. The
$(p,2p)$ reaction is subject to large attenuation, but a high
resolution experiment picking protons from $s$-orbits, for example, can probe
densities that are of the order of 30\% of $\rho _0$.

This work was supported by the Fund for Scientific
Research Flanders and the Research Board of Ghent University.

%\bibliography{QFSPRL}

\end{document}